# Reset dynamics and latching in niobium superconducting nanowire single-photon detectors

**Authors**

Anthony J. Annunziata,[1] Orlando Quaranta,[2,3] Daniel F. Santavicca,[1] Alessandro Casaburi,[2,3] Luigi Frunzio,[1,2] Mikkel Ejrnaes,[2] Michael J. Rooks,[1] Roberto Cristiano,[2] Sergio Pagano,[2,3] Aviad Frydman,[4] Daniel E. Prober[1]

[1]Dept. of Applied Physics, Yale University, New Haven, CT 06511

[2]CNR-Istituto di Cibernetica "E. Caianiello," Pozzuoli, Italy 80078

[3]Dip. di Fisica "E. R. Caianiello," Università di Salerno, Baronissi, Italy 84081

[4]Dept. of Physics, Bar Ilan University, Ramat Gan, Israel 52520

**Abstract**

We study the reset dynamics of niobium (Nb) superconducting nanowire single-photon detectors (SNSPDs) using experimental measurements and numerical simulations. The numerical simulations of the detection dynamics agree well with experimental measurements, using independently determined parameters in the simulations. We find that if the photon-induced hotspot cools too slowly, the device will latch into a dc resistive state. To avoid latching, the time for the hotspot to cool must be short compared to the inductive time constant that governs the resetting of the current in the device after hotspot formation. From simulations of the energy relaxation process, we find that the hotspot cooling time is determined primarily by the temperature-dependent electron-phonon inelastic time. Latching prevents reset and precludes subsequent photon detection. Fast resetting to the superconducting state is therefore essential, and we demonstrate experimentally how this is achieved.

## I. Introduction

Superconducting nanowire single-photon detectors (SNSPDs) offer high detection efficiency for visible and near infrared photons, with high count rates, very small timing jitter, and low dark count rates.[1-5] Typical detectors consist of a current-biased superconducting niobium (Nb), niobium nitride (NbN), or niobium titanium nitride (NbTiN) nanowire patterned into a meander, as seen in Figure 1(a). In this meander geometry, the nanowire length is proportional to the detection area. SNSPDs are particularly useful in applications that require high-count-rate single-photon detection in the near infrared, such as photon-counting communication[6] and quantum key distribution.[7] Development of NbN SNSPDs is most advanced. Although NbN SNSPDs offer higher count rates than most other near infrared single photon detectors,[8] the count rate in large-area meander devices is limited by the kinetic inductance of the nanowire, which is proportional to the nanowire length. For a small area detector (short nanowire), the count rate can be higher, but very short nanowires will latch into a finite voltage state instead of self-resetting to the superconducting state after detecting a photon. Latching precludes practical use of a small-area SNSPD at high count rate.

The count rate in a properly resetting SNSPD is set by the electrical time constant $\tau_r = L_K/R_L$, where $L_K$ is the kinetic inductance of the nanowire, proportional to its length, and $R_L$ is the load resistance of the readout circuit.[9] Before a photon is absorbed, the nanowire is superconducting with dc bias current, $I_b$, that is less than the critical current at the base temperature, $I_{co}$. An absorbed photon will create a localized hotspot in the nanowire, which has a finite resistance, $R_d$, that drives the bias current into the load, $R_L$. The equivalent circuit for a device with a finite resistance hotspot is seen in Figure 1(b). In the desired mode of operation, the SNSPD will self-reset to the superconducting state. This occurs because the hotspot cools quickly after the current is shunted out of the nanowire, so that $R_d$ abruptly returns to zero. The device current, $I_d(t)$, then exponentially returns to its initial value before photon absorption, $I_b$, with a time constant of $\tau_r$. The equivalent circuit for this current return stage is shown in Figure 1(c). Full reset requires a time of approximately $3\,\tau_r$ so that $I_d(t)$ will be restored to approximately 95% of $I_b$. This is necessary in order to have a high probability of detecting the next photon that is absorbed in the nanowire.[10] The maximum count rate is therefore the reciprocal of this full reset time, $\approx (3\,\tau_r)^{-1}$. The reset time can be reduced by increasing $R_L$[10, 11] or decreasing $L_K$[12-15]. If $\tau_r$ is too small, however, the device will not self-reset but instead will latch into a finite voltage state where it is not sensitive to photons.[10, 11, 16]

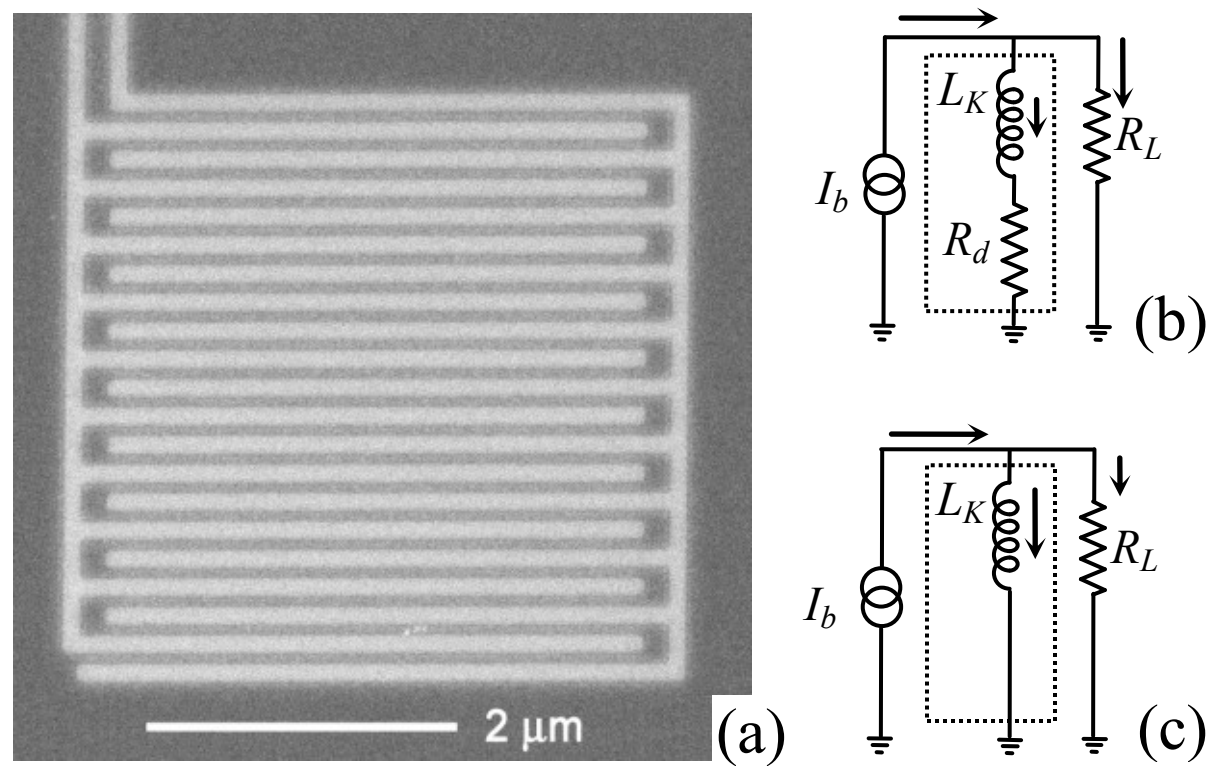

FIG. 1: (a) scanning electron micrograph of a meandered Nb SNSPD with a detection area of 16 μm$^2$. The dark regions are Nb, while the sapphire substrate is lighter; (b) equivalent circuit for the nanowire and readout with a photon-induced hotspot resistance, $R_d$. Prior to photon absorption, $R_d = 0$, and $I_b$ flows fully through $L_K$. The nanowire is outlined by the dotted box; the arrows indicate the flow of current. (c) equivalent circuit as the bias current is returning to the device after the hotspot has cooled back to the zero-resistance state. For this stage, the inductive time constant of this circuit, $L_K/R_L$, sets the time-scale for reset of the detector, which characterizes the decay of the load current toward zero.

In this article, we study the reset and latching dynamics of Nb SNSPDs using experiments and numerical simulations. We compare these results to our measurements of NbN SNSPDs, and to those of other groups.[1-4, 9-11] Reset and latching have been studied in NbN SNSPDs.[11] The present study is the first to examine the reset and latching dynamics in Nb SNSPDs.[11] A Nb SNSPD has less kinetic inductance than a NbN SNSPD of the same geometry, however, Nb has a longer electron-phonon time than NbN. As will be shown, the kinetic inductance and the electron-phonon time both play a significant role in determining the reset dynamics, which makes this comparison instructive. This work should be relevant to understanding the limits on the count rates of SNSPDs made from other materials[5, 15, 16] The present work also explains our previously reported experimental results for Nb SNSPDs.[17,18]

## II. Background

Models for the operation of NbN SNSPDs have been presented.[10,11,19–21] Ref. 11 studied the phenomenon of latching in detail. The analysis in Ref. 11 uses a phenomenological model of the heating and cooling of the photon-created resistive hotspot to determine if a NbN SNSPD will latch. According to Ref. 11, electrothermal feedback creates a situation where a resistive hotspot is either stable or unstable in the steady state, depending on $I_b$ and $\tau_r$. Formation of a stable hotspot, known as latching, precludes operation of the SNSPD. Thus, in the case of the SNSPD, parameters that give "unstable" hotspot behavior are desired.

Solutions to the model in Ref. 11 were obtained analytically by determining the stability of the hotspot under small sinusoidal perturbations. This type of small-signal stability analysis does not model the time-dependent formation, evolution, and decay of the hotspot. The predictions of the model in Ref. 11 were fit to data from NbN SNSPDs. By varying several of the phenomenological parameters of the model, good agreement between the model predictions and experimental data was found. However, some of the phenomenological parameters used in Ref. 11, notably the "hotspot temperature stabilization time," are not directly connected to the microscopic physical processes, such as electron-phonon scattering, phonon-escape and electron diffusion. These physical processes govern energy relaxation in superconducting thin films, and are important for understanding other non-equilibrium superconducting devices such as hot electron bolometers and transition edge sensors.[22, 23] Identifying which of these physical processes plays the key role in the SNSPD is important. Additionally, in Ref. 11, the values of some parameters that are obtained by the fitting procedure appear to be different from those of independent measurements.[16-18, 22-27] For example, the results in Ref. 11 based on fitting imply a value of the thermal conductivity, κ, in NbN of approximately 0.0017 W/K-m. Direct measurements of the thermal conductivity in NbN, and calculations based on the Wiedemann-Franz law, have obtained κ ≈ 0.16 W/K-m.[10] We do believe that the model in Ref. 11 provides guidance that is useful in understanding trends within the data presented, and also may capture physical effects occurring at the superconducting-normal interface of the hotspot that may be particularly important to understand latching in NbN SNSPDs, with their higher resistivity and shorter length normal-superconducting interfaces.

In the present work, we develop a model to analyze latching in Nb SNSPDs based on microscopic physical processes. These are well-known to govern thermal relaxation in superconducting thin films and nanowires. We find that a photon-created hotspot can stabilize to either a finite resistance or cool back to zero resistance, depending on $I_b$, $L_K$, and $R_L$. The factors that determine whether or not a hotspot will latch are identified by examining the full dynamics of the hotspot formation and how the dynamics depend on $I_b$, $L_K$, and $R_L$. We find that almost all of the energy stored in the kinetic inductance of the nanowire, $\frac{1}{2}L_K I_b^2$, is dissipated into the hotspot as Joule heat when the hotspot forms. This inductive energy dissipation occurs over a time scale that is significantly less than 1 ns in most Nb devices. We conclude that this total inductive energy determines whether a device will latch for a specific value of the current return time, $\tau_r = L_K/R_L$. The predictions of our model agree well with measurements of Nb SNSPDs with no free parameters. The parameters we use in our model are based on independent measurements.

We find that Nb SNSPDs are significantly more susceptible to latching than NbN devices because cooling by phonon emission in Nb is much slower than cooling by phonon emission in NbN. Although this slower cooling is less desirable for high count rate operation, it makes Nb a good material in which to study latching. The conclusions we draw from our model of Nb SNSPDs should be applicable to SNSPDs fabricated from other materials as well. We also tested NbN devices to verify that our measurement methods did not introduce any spurious behavior in our measurements of Nb SNSPDs. Proper shielding, filtering, and clean, low noise microwave design are critical to these measurements. Our measured results for NbN are similar to those reported in Ref. 11, confirming the validity of our measurements.

## III. Devices Tested

The fabrication procedures for the Nb and NbN devices measured in this work are described in Refs. 18 and 24, respectively. All Nb nanowires were ≈ 7.5 nm thick, 100 nm wide and approximately 110 Ω/square just above the critical temperature, $T_c$. All NbN nanowires were ≈ 5 nm thick, 130 nm wide and approximately 900 Ω/square just above $T_c$. The devices studied had good (≈ 5%) detection efficiency for 470 nm photons and uniform line width; no significant constrictions were apparent.[25] The detection efficiency of these devices can likely be improved significantly by using optical structures designed to increase the absorption in the Nb film, as was done for NbN SNSPDs in Ref. 4. A summary of the devices tested is given in Table 1. The variation in $T_c$, and therefore of $I_{co}$, for the Nb devices appears to be due to variation in thickness between samples, which affects the superconductivity strongly because the devices are very thin.[15] The measurement setup is described in Ref. 18 and an equivalent circuit is given in Figure 2(a). A key feature of the experimental setup is a set of cryogenic, remote controlled RF switches. These enable measurement during one cooldown of a given device with various shunt resistors in parallel with the transmission line but close to the device. This gives total load resistances $R_L$ = 50 (with no shunt), 33, 25 (with a 50 Ω shunt), 17, or 15 Ω. We measured $L_K$ independently by incorporating each device into a resonant circuit with known capacitance and measuring the resonant frequency for a range of temperatures below $T_c$ and for a range of currents below $I_{co}$ using a network analyzer. The measured values at 1.7 K are reported in Table 1 for $I_b$ = 0. The dependence of $L_K$ on $I_b$ is weak, varying by ≈ 5% in Nb nanowires as $I_b$ is increased to just below $I_c$. This will be discussed separately in a future publication.

| **Parameter** | **Nb (A)**<br>10 x 10 μm$^2$ | **Nb (B)**<br>4 x 4 μm$^2$ | **Nb (C)**<br>10 μm line | **NbN (D)**<br>5 x 5 μm$^2$ | **NbN (E)**<br>5 μm line |
|---|---|---|---|---|---|
| length: $L$ | 500 μm | 80 μm | 10 μm | 105 μm | 5 μm |
| $T_c$ | 4.5 K | 3.9 K | 4.0 K | 10 K | 10 K |
| $I_c$ (1.7 K) | 8.2 μA | 6.2 μA | 6.4 μA | 26.2 μA | 25.4 μA |
| $L_K$(1.7 K) | 235 nH | 60nH | 15 nH | 120 nH | 16 nH |

Table 1: Parameters of the devices studied in this work. The reported values of the kinetic inductance for each device in the table include approximately 10 nH of inductance from the measurement leads.[28] The sheet resistances are approximately 110 Ω/square and 900 Ω/square for Nb and NbN, respectively.

## IV. Model

The simulations we have performed are based on numerical solutions to a two-temperature model of heat flow similar to the model used to analyze NbN SNSPDs in Ref. 20. In two-dimensions, the governing equations of the model are:

$$\dot{T}_e \qquad \frac{1}{\tau_{e-ph}(T_e)}\left(T_e - T_{ph}\right) + \frac{1}{C_e(T_e)} j_d(x,y,t)^2 \rho_d(x,y,t) + D_e \nabla^2 T_e, \tag{1}$$

$$\dot{T}_{ph} \qquad \frac{1}{\tau_{e-ph}(T_e)} \frac{C_e(T_e)}{C_{ph}(T_{ph})}\left(T_e - T_{ph}\right) - \frac{1}{\tau_{esc}}\left(T_{ph} - T_o\right) + D_{ph} \nabla^2 T_{ph}, \tag{2}$$

where $T_e = T_e(x,y,t)$ and $T_{ph} = T_{ph}(x,y,t)$, are the electron and phonon temperatures, $T_o$ is the substrate temperature, $\tau_{e-ph}(T_e)$ is the electron-phonon inelastic scattering time, $\tau_{es}$ is the escape time for phonons (equal to 40 ps in all simulations)[23] , $D_e$ and $D_{ph}$ are the electron and phonon diffusivities, $C_e(T_e)$ and $C_{ph}(T_{ph})$ are the heat capacities of the electrons and phonons per unit volume, $j_d(x,y,t)$ is the current density, and $\rho_d(x,y,t)$ is the resistivity. In all simulations of Nb SNSPDs, we use $\tau_{e-ph}(T_e) = \tau_{e-ph}(6.5\ \mathrm{K})(6.5\ \mathrm{K}/T_e)^2$ with $\tau_{e-ph}(6.5\ \mathrm{K}) =$ 2.0 ns,[23] $D_e = 1\ \mathrm{cm^2/s}$,[27] $D_{ph} = 0.1\ \mathrm{cm^2/s}$, $C_{ph}(T_{ph}) = 9.8\times10^{-6} T_{ph}^3$ (J*cm$^{-3}$K$^{-1}$).[10, 29] For the electronic heat capacity, in the normal state we use $C_{eN}(T_e) =$ $5.8\times10^{-2}\ T_e$ (J*cm$^{-3}$K$^{-1}$),[27] and in the superconducting state we use $C_{eS}(T_e) = 0.92*exp(-\Delta(T_e)/k_B T_e)$ (J*cm$^{-3}$K$^{-1}$),[27-30] where $\Delta(T_e) = 1.76 k_B T_c*(1-T_e/T_c)^{1/2}$.[29]

The total current flowing through the device, $I_d(t)$, is determined by the readout circuit (Figure 2(a)) and therefore obeys the equation:

$$\dot{I}_d(t) \qquad \frac{1}{L_K}\left(R_d(t) + R_L\right) I_d(t) + R_L \frac{I_b}{L_K}, \tag{3}$$

which is obtained using Kirchhoff's laws, where $R_d(t)$ is the resistance of the nanowire and where $I_b = V_b/R_b$ in Figure 2(a). The spatial distribution of the current density, $j_d(x,y,t)$, is determined by the spatially-dependent resistivity of the device, $\rho_d(x,y,t)$. If the coordinate system is oriented such that the positive x-axis is along the length of the nanowire in the direction of current flow, the total device current at position ($x$) for a wire of width $w$, and thickness $d$ much smaller than the magnetic penetration depth, is given by:

$$I_d(t) = I_d(x,t) = d\int_0^w j_d(x,y,t)dy\,, \tag{4}$$

which, by conservation of charge, must be equal at all points ($x$). The local resistivity, $\rho_d(x,y,t)$, will depend on the whether the point ($x,y$) in the material is in the superconducting or normal conducting state. In our model, the resistivity is defined by:

$$\rho_d(x,y,t) = \rho_o\left[1-\left[H\left(T_c - T_e\left(x,y,t\right)\right)\cdot H\left(j_c - j_d\left(x,y,t\right)\right)\right]\right], \tag{5}$$

where $H$ is the Heaviside step function and $\rho_o$ is the normal state resistivity of the film. Thus, $\rho_d(x,y,t)$ is equal to zero or $\rho_o$, depending on temperature and current density. Since only those sections of the strip at point ($x$) that are normal for all values of ($y$) at ($x$) will contribute to $R_d$, it follows that for a wire of width $w$,

$$R_d(t) = \frac{l_{norm}(t)}{dw}\rho_o\,, \tag{6}$$

where the normal length, $l_{norm}(t)$, is calculated numerically and is the length over which the resistive hotspot occupies the entire width of the wire. Typically, the maximum value of $l_{norm}$ is much less than the total nanowire length, $l$. Finally, we calculate the effective critical current of the device as a function of time, $I_c(t)$. The effective critical current is defined as the minimum critical current along the length of the nanowire using the Ginzburg-Landau expression for the temperature dependence:

$$I_c(t) = \min_x \left[ d \int_0^w j_c(T_e = 0) \left( 1 - \frac{T_e(x,y,t)_e}{T_c} \right)^{3/2} dy \right], \quad (7)$$

with $j_c(T_e = 0)$ determined for each device by equating the Ginzburg-Landau expression for $I_c(1.7\text{K}) = I_c(T_e = 0)(1-1.7/T_c)^{3/2}$ to measurements of $I_c(1.7\text{ K})$ and setting $j_c = I_c/wd$.[29] As defined, the effective critical current is assumed to be independent of $I_d$. Since the effective critical current is only determined by temperature, once the hotspot begins to cool below the critical temperature, $I_c(t)$ becomes a measure of the thermal relaxation of the highest temperature region of the hotspot. Thus, the time scale over which the critical current returns to near its equilibrium value is the hotspot thermal relaxation time, $\tau_c$. This is the time required for hot electrons to return to near their equilibrium temperature, $T_o$.

We have implemented a numerical solution to these equations using MATLAB.[31] The device is represented by a two dimensional grid with longitudinal grid spacing $\Delta x$ and transverse grid spacing $\Delta y$. At each grid point, the electron and phonon temperatures are defined. From these temperatures, all temperature-dependent quantities are defined. When a volume-dependent quantity such as the heat capacity is calculated, it is calculated over the volume of the cell centered at the grid point $(x,y)$ where the volume of the cell is equal to $d\Delta x\Delta y$ with typical grid spacing of $\Delta x = \Delta y = 5$ nm. The absorption of a photon is simulated by increasing the temperature of one grid point in one time step $\Delta t$ such that

$$T_e(x_o, y_o) = T_o + \frac{hf}{C_e(T_o)\cdot(\Delta x \Delta y d)}, \quad (8)$$

where $(x_o,y_o)$ is the grid point where the photon is absorbed, $h$ is Planck's constant, and $f$ is the frequency of the photon.

## V. Results

The detection cycle in an SNSPD, as illustrated by the simulations, can be divided into three distinct stages, labeled on the lower curve in Figure 2(b): (i) The nanowire is biased in the superconducting state with a dc bias current, $I_b$, below the critical current at the base temperature, $I_{co} = I_c(T_o)$; here, $R_d = 0$. (ii) A photon creates a resistive hotspot whose resistance, $R_d(t)$, increases quickly due to the fast dissipation of the inductive energy stored in $L_K$. As a result, most of $I_b$ is shunted into $R_L$, which has much lower resistance than $R_d$; (iii) In a self-resetting device, after most of the bias current has transferred to $R_L$, the hotspot quickly returns to the zero-resistance state and the current slowly begins to transfer back into the device with a return time constant $\tau_r = R_L/L_K$. In a latching device, stages (i) and (ii) are identical to those in a self-resetting device, however stage (iii) does not occur and $R_d$ remains finite. Latching is seen in Figure 2(b) for $I_b = 8.1$ μA. As will be shown next, latching occurs in this device ($L_K = 235$ nH, $R_L = 50\ \Omega$) at the larger value of $I_b$ because of the greater heating that occurs due to the larger amount of stored inductive energy, $\approx \frac{1}{2}L_K I_b^2$, that is dissipated into the hotspot. If the hotspot does not cool quickly enough as the current begins to return to the device, the device will remain in (latch into) the resistive state. The measured latching pulse in Figure 2(b) is for a single shot measurement. It has more noise than the measured self-resetting pulse because the self-resetting waveform displayed is an average of many pulses. The measured slow decay of $I_L(t)$ in the latching case for $t > 5$ ns is due to the ac coupling of the amplifier. This slow decay is not observed in the simulation, which assume a dc-coupled amplifier.

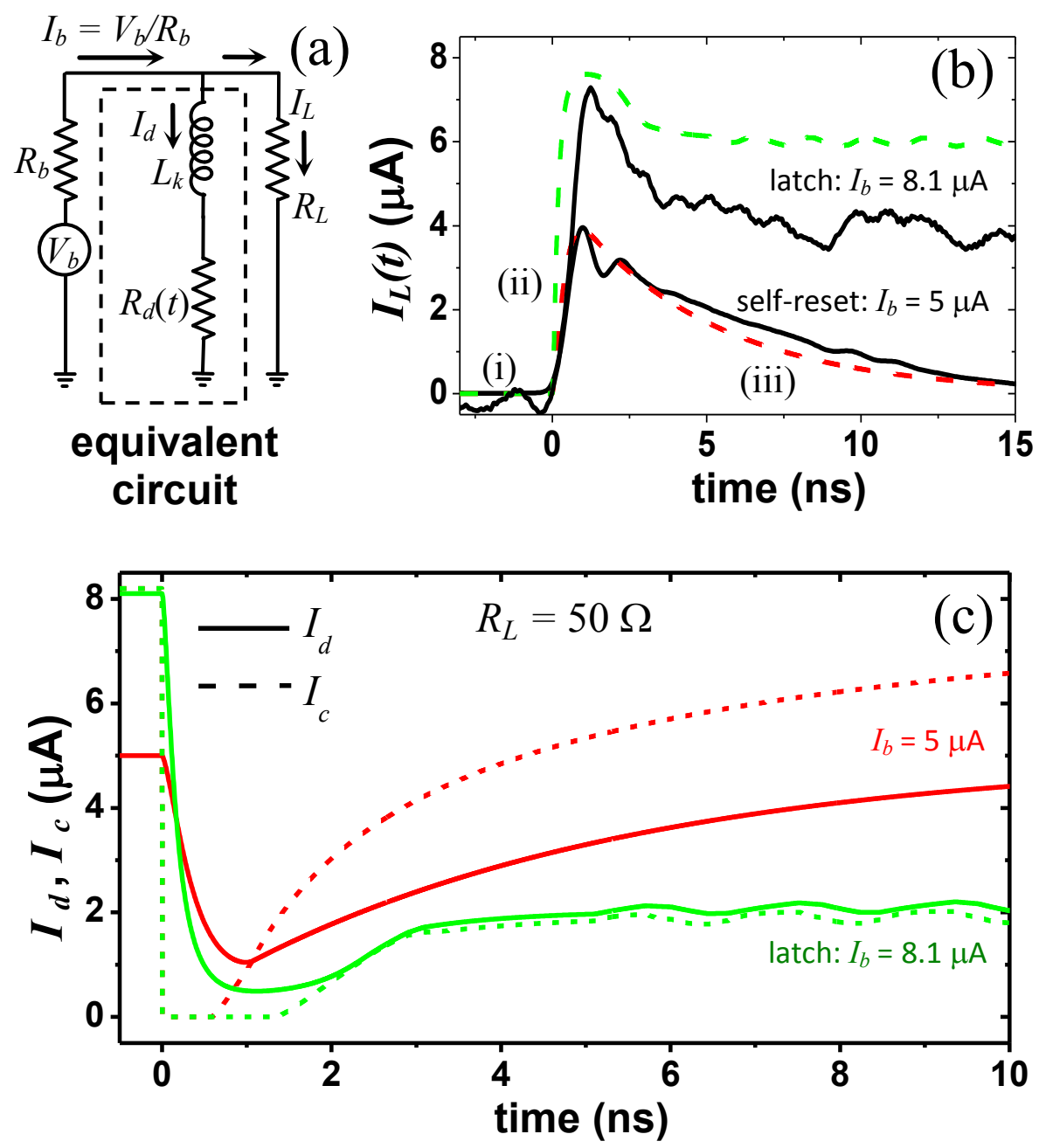

FIG. 2 (color online): (a) equivalent circuit; $R_b >> R_L$, $R_d$; (b) measured (solid lines) and simulated (dashed lines) output pulses, $I_L(t)$, with the self-resetting case labeled by the three regimes of operation: (i) the device is in equilibrium with $R_d = 0$; (ii) initial heating: a photon has been absorbed, the hotspot is growing and the current is transferring into the load; (iii) the hotspot resistance has returned to zero, and the current is returning to the device with a time constant $\tau_r = L_K/R_L$; (c) numerical results for $I_c(t)$ (dashed lines) and $I_d(t) = I_b - I_L(t)$ (solid lines) for $I_b$ = 5.0 (red) and 8.1 μA (green); the latter shows the latching case. We find $I_{latch}$ = 7.1 μA from these simulations. We note that $I_c$ as defined here can be finite even with $R_d > 0$ because $I_c$ is calculated according to equation (7) and therefore only depends on the electron temperature, and not on $I_d$. $R_d$ is calculated from the simulation according to equation (6).

It is essential that $I_b$ be close to $I_{co}$ for high detection efficiency.[20] Thus, self-reset for $I_b \lesssim I_c$ is necessary for a device to have both a high count rate and high detection efficiency. In Figure 2(c), we model the effect of increasing $I_b$ from 5 μA to 8.1 μA by computing both the effective time-dependent critical current, $I_c(t)$, and the device current, $I_d(t)$, as functions of time. The fast dissipation of the inductive energy, $\frac{1}{2}L_K I_b^2$, that occurs after the photon is absorbed drives $I_c$ quickly to zero (Figure 2(c)) and $R_d$ to a large value. This large value of $R_d$ drives $I_d$ to near zero because $I_b$ is shunted almost entirely into the lower resistance load, $R_L$. This entire process occurs in a very short time, < 1 ns in most devices, as observed from the rise time of the current pulses in Figure 2(b). The cooling time in Nb, as we will show, is much longer than the sub-nanosecond time to dissipate the inductive energy, $\frac{1}{2}L_K I_b^2$.

Self-reset occurs when the trajectories of $I_c(t)$ and $I_d(t)$ are such that $I_d(t) < I_c(t)$ as the hotspot cools**.** For device (A) with $R_L = 50\ \Omega$, Figure 2(c) shows that self-reset occurs for $I_b = 5$ μA but not for $I_b = 8.1$ μA. When $I_b = 8.1$ μA, $I_d(t)$ remains above $I_c(t)$ as the hotspot begins to cool, maintaining a finite hotspot resistance. The resulting dc Joule heating (occurring after the initial energy dissipation) enforces a stable, finite resistance hotspot state. The resistance of this stable state may be less than the peak resistance of the hotspot, which occurs just after the fast initial dissipation of the inductive energy and before any substantial cooling has taken place. Note in Figure 2(c) for $I_b = 8.1$ μA that the finite value of $I_c$ for long times indicates that the steady-state temperature at the center of the hotspot is less than $T_c$; nevertheless, the hotspot remains resistive because $I_d > I_c$. We define a latching current, $I_{latch}$, as the lowest value of $I_b$ that results in latching. For device (A) in Figure (1) with $R_L = 50\ \Omega$, the model predicts that $I_{latch} =$ 7.1 μA.

We now show that our numerical simulations agree with experimental measurements. In Figure 3 we plot the normalized latching current measured for several Nb SNSPDs versus $\tau_r = L_k/R_L$. We also plot predictions for the latching current obtained by simulating devices with the same parameters as those measured. In addition to the measurements and simulations for Nb devices, we also plot experimental measurements of the latching current in NbN devices, to provide a comparison of the two different materials. Each device has a different value of $L_K$ (see Table 1). We vary $R_L$ to change $\tau_r$ in both simulations and experiments. For Nb device (A), we plot predictions for two values of $D_e$ (1.0 cm$^2$/s and 0.25 cm$^2$/s) to show that faster diffusion of hot electrons (larger $D_e$) makes a device less susceptible to latching. A value of $D_e$ = 1.0 cm$^2$/s was used for all other Nb simulations. The simulation predictions for Nb SNSPDs and the experimental measurements of Nb SNSPDs are in approximate quantitative agreement. In the simulations, there are no free fitting parameters used. We believe the discrepancies between simulated and measured values of $I_{latch}$ are primarily due to uncertainty in the values of the material parameters used as inputs to the model.[10, 23, 27-29]

We find in experiments that a Nb SNSPD is much more sensitive to latching than a NbN SNSPD with the same area and value of $R_L$. Still, we observe that NbN SNSPDs are also susceptible to latching, as discussed in Ref. 11. A trend we observe for both Nb and NbN devices is that with larger kinetic inductance, a larger value of $\tau_r$ is required to achieve the same normalized value of $I_{latch}$. For larger $L_K$, the dissipated energy ≈ $\frac{1}{2} L_K I_b^2$ is larger. For example, in Figure 3 the circled data points for devices (A) and (B) are both at $I_{latch} = 0.95 I_c$; device (A) has $L_K$ = 235 nH, $R_L$ = 25 Ω, and $\tau_r$ = 9.4 ns while device (B) has $L_K$ = 60 nH, $R_L$ = 17 Ω, and $\tau_r$ = 3.5 ns.

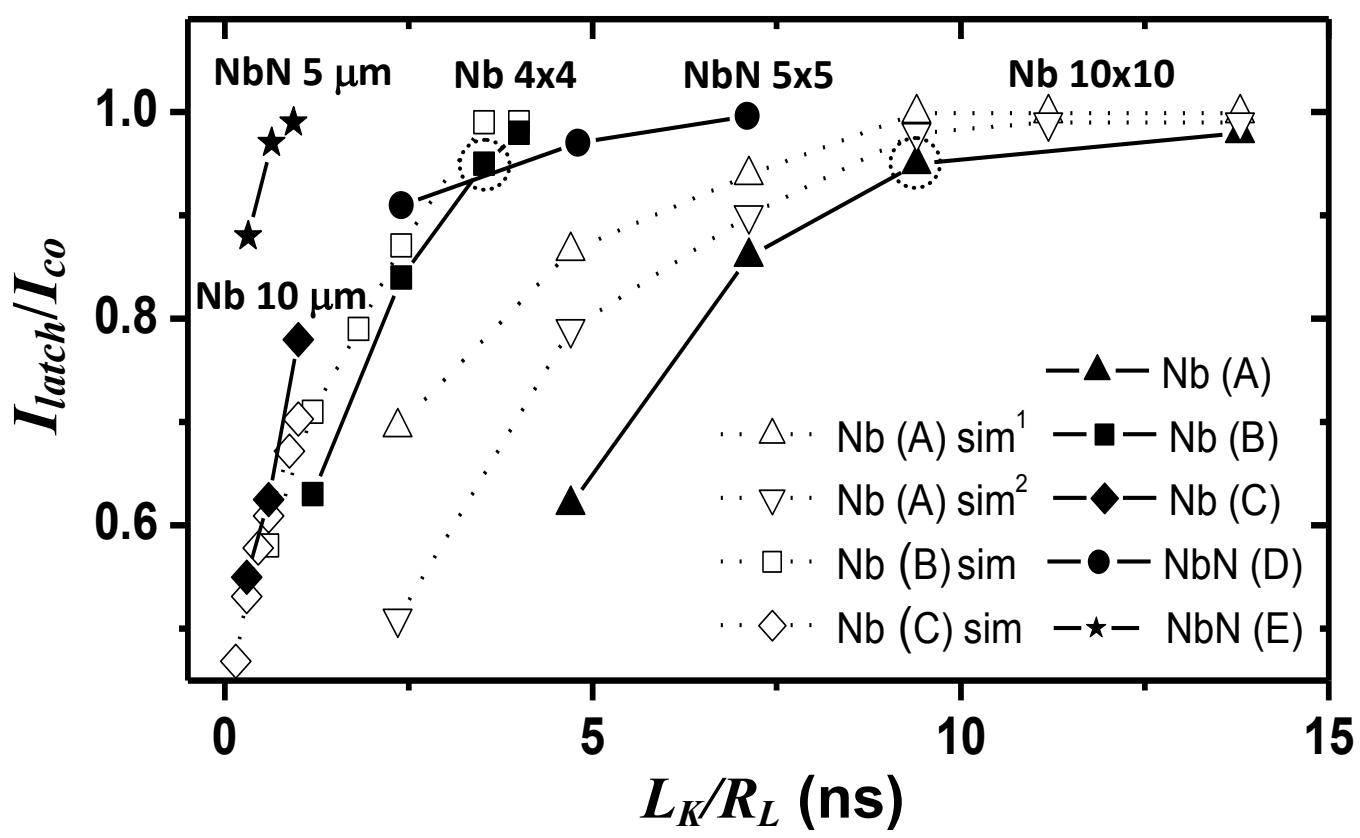

FIG 3: Normalized latching current predicted by the model (open symbols) and measured (solid symbols) versus $\tau_r = L_K/R_L$, obtained by varying $R_L$ in the simulations and experiments. For the 10x10 μm$^2$ meander, we plot predictions for two values of the diffusion constant: $^1D_e = 1$ cm$^2$/s, as used in all other simulations, and $^2D_e = 0.25$ cm$^2$/s, which is used only here.

From simulations, we find that latching occurs when the current return time, $\tau_r$, is significantly shorter than the hotspot cooling time, $\tau_c$. We define $\tau_c$ as the time interval between when $I_c(t)$ first becomes finite and when $I_c(t) = 0.63I_{co}$ (see Figure 3). This criterion allows us to compare $\tau_r$ to $\tau_c$. Simulations show that $\tau_c$ depends on material parameters (e.g. the electron-phonon time) as well as on the inductive energy $\frac{1}{2}L_K I_b^2$. After photon absorption and the fast dissipation of the inductive energy, the hotspot temperature is significantly greater than the equilibrium temperature, $T_o$. The maximum temperature is largest for the largest stored energy. We find from simulations of Nb SNSPDs that the temperature-dependent electron-phonon interaction time in Nb is the dominant component of $\tau_c$, but not the only component. For small temperature excursions, $\tau_c$ would be equal to the electron-phonon time,[32] $\tau_{e\text{-}ph}(T_e) \sim (T_e)^{-2}$.

However, for the large temperature changes that occur in an SNSPD, the thermal relaxation has a more complex behavior than simple exponential relaxation and also depends on $R_d(t)$ and on the out diffusion of hot electrons. Thus, simulations are needed to predict the exact trajectories of $I_c(t)$ and $I_d(t)$, and therefore if a device will latch or self-reset.

Simulations confirm that the inductive energy dissipation in the hotspot occurs on a time scale much shorter than $\tau_c$ and can be much larger than the energy of the photon that initiates the hotspot. From simulations, we find that the exact energy dissipated in the hotspot as it is forming is equal to = $\frac{1}{2}L_K(I_b^2-I_{min}^2) + hf$, where $I_{min}$ is the minimum value of $I_d$ and $hf$ is the energy of the absorbed photon, where $h$ is Planck's constant and $f$ is the frequency of the photon. For most practical devices (large area and therefore large $L_K$), $hf << \frac{1}{2}L_KI_b^2$. In all measurements and simulations, we find that $I_{min} << I_b$, so that the total initial dissipated energy is approximately equal to the inductive energy, $\frac{1}{2}L_KI_b^2$ for all but the shortest nanowires, e.g., device (C), where the photon energy, $hf$, is significant. In a self-resetting device, the total amount of energy dissipated in a detection event is this initially dissipated energy, $\frac{1}{2}L_KI_b^2$, since the hotspot quickly cools back to near $T_o$ and $R_d = 0$ after the inductive energy is dissipated. In a latching device, $I_d(t)$ remains above $I_c(t)$ as the hotspot begins to cool. In this case, the hotspot remains resistive as $I_d$ begins to increase, causing additional Joule heating after the inductive energy is dissipated. This additional Joule heating enforces a stable dc hotspot resistance.

In all devices, a larger value of $\frac{1}{2}L_KI_b^2$ leads to a hotspot with a greater peak temperature. This increases the cooling time, $\tau_c$, since more energy must be transferred to the phonon system and ultimately to the substrate in order for the hotspot to cool to near $T_o$. Since $\frac{1}{2}L_KI_b^2$ and $\tau_c$ are both independent of $R_L$, we can reduce $R_L$ in order to increase $\tau_r$ without affecting the cooling time, $\tau_c$. We can therefore eliminate latching in any device by sufficiently reducing $R_L$. However, reducing $R_L$ decreases the count rate, proportional to $(3\tau_r)^{-1}$ and reduces the output signal, $\approx I_b R_L$.

In Figure 4, we show the effects of changing $L_K$ and $I_b$ for four different situations. We plot simulations of $R_d(t)$ and $I_c(t)$ for Nb devices with $R_L = 25\ \Omega$. (Note that in Figure 2, $R_L = 50\ \Omega$ was used.) The inductive energy is varied by changing $L_K$ (compare device B with $L_K = 60$ nH to device C with $L_K = 15$ nH) and by changing $I_b$ (compare device (A) with $I_b = 5\ \mu$A to device (A) with $I_b = 8.19\ \mu$A). The plots of the simulated data clearly show that the peak value of $R_d$ as well as the value of the cooling time, $\tau_c$, increases for larger values of $\frac{1}{2}L_K I_b^2$. For example, in Figure 4, $\tau_c$ increases from 2.2 to 5.0 ns when the value of $\frac{1}{2}L_K I_b^2$ is increased from 5 eV for device (C) to 85 eV for device (A) with $I_b = 8.19\ \mu$A; $hf = 2.6$ eV in these simulations.

In addition to circuit parameters, the thermal relaxation process also depends on material parameters. In particular, the cooling time, $\tau_c$, decreases when the electron-phonon time is decreased. Also, if $D_e$ is increased, $\tau_c$ decreases due to faster out diffusion of hot electrons (see device (A) in Figure 3 where two values of $D_e$ are considered). The dependence of $\tau_c$ on these material parameters likely explains the difference in the observed latching behavior between Nb and NbN SNSPDs (Figure 3). NbN has a much shorter electron-phonon time than Nb at all relevant temperatures.[26] This gives less latching because it more than compensates for the slower diffusion in NbN than in Nb and for the larger value of $\frac{1}{2}L_K I_b^2$ in NbN SNSPDs compared to Nb SNSPDs of similar geometry.

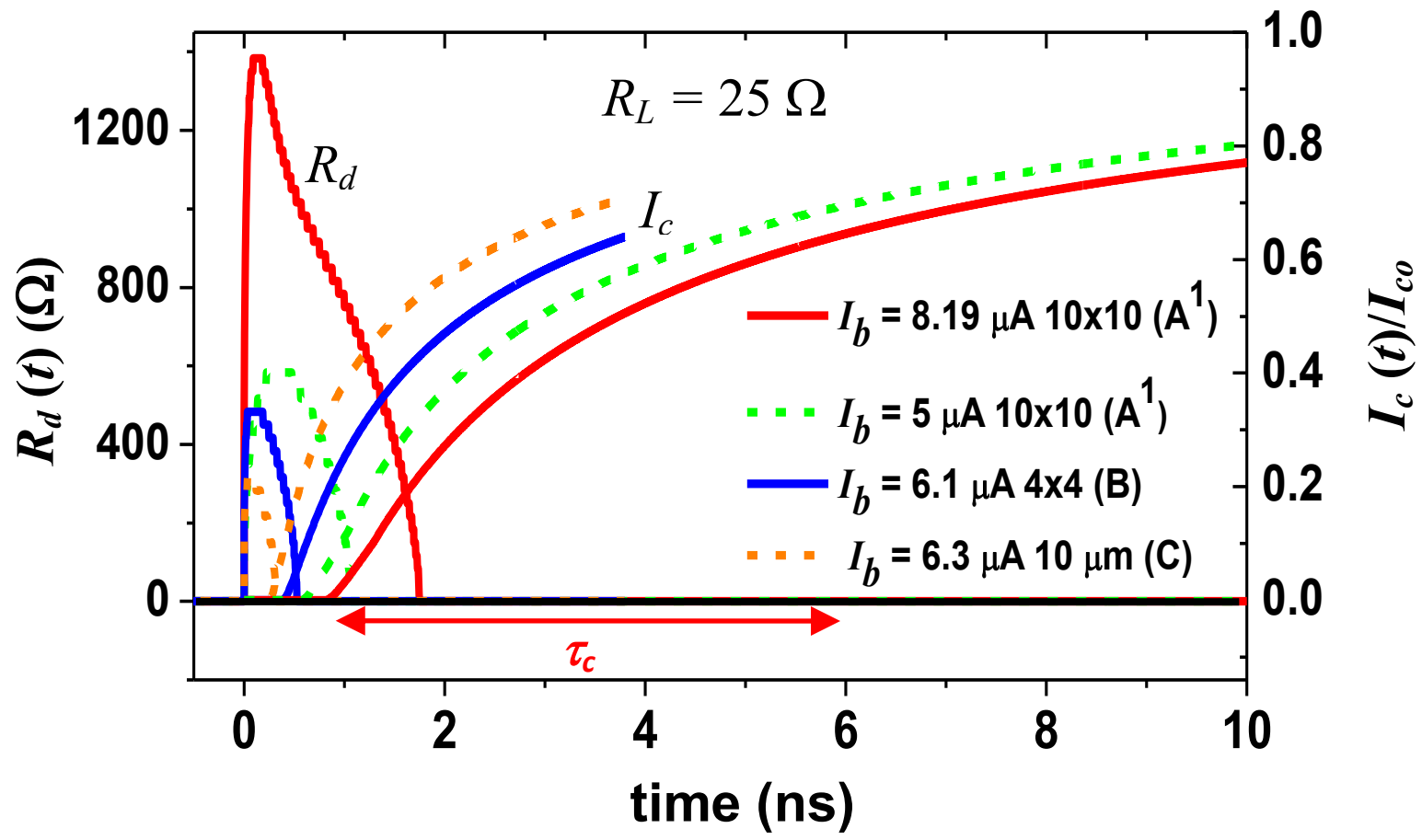

FIG. 4 (color online): Simulation results for Nb devices (A), (B) and (C) showing the dependence of $R_d(t)$ and $I_c(t)$ on $I_b$ and $L_K$; $R_L$ = 25 Ω; $\tau_c$ is labeled for device (A) with $I_b$ = 8.19 μA and is equal to 5.0 ns. For device (C) $\tau_c$ = 2.2 ns (not labeled).

## VI. Conclusion

We have demonstrated that our microscopic model of the electron heating and cooling can predict the observed reset and latching behavior in a Nb SNSPD. Our simulations and measurements of Nb SNSPDs show behavior qualitatively similar to our measurements of NbN SNSPDs and to the results reported in Ref. 11 for NbN SNSPDs. In Nb SNSPDs, we find that the fast initial heating caused by the Joule dissipation of the energy stored in the kinetic inductance of the nanowire determines whether a Nb device will latch or self-reset. In Nb, latching occurs over a wider range of operating parameters than for NbN SNSPDs. This is due to the much larger electron-phonon time in Nb. Our model suggests that the maximum count rate for a Nb SNSPD with small $L_K$ (e.g., device (C)) is $(3\tau_{c,min})^{-1} \approx 150$ MHz. To achieve this count rate with device (C), however, a value of $R_L \approx 7$ Ω would be required to prevent latching if $I_b$ is just below $I_c$ (see Figure 3).

We find that the intrinsic timescale for cooling of the hot electrons is the electron-phonon inelastic time. This sets the lower limit to the reset time of an SNSPD. Thus, reducing the kinetic inductance helps only insofar as it reduces the inductive energy that is dissipated as the hotspot forms, which reduces the cooling time. Based on this conclusion, a NbTiN SNSPD, with a smaller kinetic inductance than a NbN SNSPD of the same geometry,[15] may have some advantages over NbN SNSPDs if the electron-phonon inelastic times are similar. However, even if an SNSPD were fabricated from a superconducting material with a shorter electron-phonon time than NbN, such as $MgB_2$, the phonon escape time, $\tau_{es}$, probably cannot be reduced significantly from its value for ultra thin NbN, where $\tau_{es} \approx 40$ ps.[33] Recently, new geometries that incorporate several nanowires in parallel have been explored.[34,35] The added degrees of freedom for the current in these parallel topologies may allow for faster reset than standard single-nanowire meander-type detectors of the same detection area.

This work was supported by NSF ECS-0622035 and an NSF Graduate Research Fellowship to one of us (AA).